\documentclass[article]{elsarticle}
\makeatletter
\def\ps@pprintTitle{%
 \let\@oddhead\@empty
 \let\@evenhead\@empty
 \def\@oddfoot{\centerline{\thepage}}%
 \let\@evenfoot\@oddfoot}
\makeatother

\expandafter\let\csname equation*\endcsname\relax

\expandafter\let\csname endequation*\endcsname\relax

\makeatletter
\let\@fnsymbol\@arabic
\makeatother

\makeatletter
\def\@makefnmark{\hbox{\@textsuperscript{\normalfont\@thefnmark}}}
\makeatother

\usepackage[hang]{footmisc}
\usepackage{setspace}
\usepackage[margin=3.5cm]{geometry}
\usepackage{rotating}
\usepackage{hyperref}
\usepackage{amsfonts}
\usepackage{amsmath}
\usepackage{bbm} 

\usepackage{slashed}
\usepackage{multirow}
\usepackage{array}
\usepackage{graphicx}
\usepackage{tikz-cd}
\usepackage{makecell}
\usepackage{adjustbox}
\usepackage{longtable}
\usepackage{setspace}
\usepackage{caption}

\usepackage[utf8]{inputenc}
\usepackage[english]{babel}
\DeclareUnicodeCharacter{BC}{}

\newtheorem{theorem}{Theorem}[section]
\newtheorem{definition}{Definition}[section]

\allowdisplaybreaks

\newcommand{\R}{\mathbb{R}}
\newcommand{\C}{\mathbb{C}}

\begin{document}

\begin{frontmatter}

\title{Invariant polynomials and machine learning}

\author{Ward Haddadin}
\address{DAMTP, University of Cambridge, Wilberforce Road, Cambridge, CB3 0WA, United Kingdom} 
\ead{w.haddadin@damtp.cam.ac.uk}

\begin{abstract}
We present an application of invariant polynomials in machine learning. Using the methods developed in previous work, we obtain two types of generators of the Lorentz- and permutation-invariant polynomials in particle momenta; minimal algebra generators and Hironaka decompositions. We discuss and prove some approximation theorems to make use of these invariant generators in machine learning algorithms in general and in neural networks specifically. By implementing these generators in neural networks applied to regression tasks, we test the improvements in performance under a wide range of hyperparameter choices and find a reduction of the loss on training data and a significant reduction of the loss on validation data. For a different approach on quantifying the performance of these neural networks, we treat the problem from a Bayesian inference perspective and employ nested sampling techniques to perform model comparison. Beyond a certain network size, we find that networks utilising Hironaka decompositions perform the best.
\end{abstract}

\end{frontmatter}

\setcounter{tocdepth}{2}

\tableofcontents

\section{Introduction}

In the era of big data, solutions to most quantitative problems only seem to be a machine learning algorithm away. Much work has been done on developing and extending machine learning tools to cope with a diverse array of applications.\footnote{The interested reader may search for ``machine learning" on-line and will instantly be bombarded with more than they can ever read about the subject.} Applying these algorithms to data has allowed their users, with varying degrees of success, to model the underlying systems of study and extract interesting, often valuable, information previously unknown to us mere humans. 

But prior to this data revolution, when one had to think analytically about the systems one was studying, generations of scientists uncovered a great deal of knowledge about physical systems and their properties. To simply ignore their work in our modern data-driven quests for knowledge seems somewhat ungrateful, and possibly wasteful. Take, for example, the setting of particle physics, where we not only know that all processes must be Lorentz-invariant, but also that they can often be permutation-invariant since the particles involved can be identical and thus indistinguishable. For examples in slightly different settings, consider physical chemistry or material science, where the properties of molecules and lattices are rotation- and translation-invariant and also potentially permutation-invariant when they contain identical atoms. These seemingly simple facts turn out to have important consequences for our attempts to model the world using data. 

In this work, we demonstrate the benefit of incorporating our knowledge of the symmetries of certain systems into the machine learning methods we use to describe them. We do this by borrowing tools from a branch of mathematics called invariant theory, which provide us with systematic methods to produce sets of variables that describe our systems whilst incorporating our knowledge of their symmetries. These methods result in generators which come in two flavours; {\em minimal algebra generators} and {\em Hironaka decompositions}. Crucially, the description via these variables loses no information about the systems which ensures that we do not miss out on any possibly interesting information in our data. Furthermore, Hironaka decompositions have the added advantage that they are redundancy-free (to be defined in Section \ref{sec:generators}), a feature which is highly desirable in any machine learning endeavour.

We choose the setting of particle physics as our playground. Here, the data we collect from collision events in colliders comes in the form of momenta, {\em i.e.} vectors, and a theorem of Weyl tells us that the generators of the Lorentz-invariant polynomials in the momenta are the dot products and the epsilon tensor contractions. By using methods constructed in previous work \cite{LPIP}, we further incorporate permutation symmetry and obtain generators of the Lorentz- and permutation-invariant polynomials in these momenta. We implement these generators in the infamous machine learning tool, neural networks (applied  to regression tasks), and test their performance compared to using the plain Weyl generators. We find a marked improvement in the training loss and a significant improvement in the validation loss on unseen data over a wide range of hyperparameters. For a different perspective on how to quantify the performance improvements, we treat these networks as Bayesian models and apply Bayesian inference methods (nested sampling) to compute the model evidences and hence perform model comparison. We find that beyond a certain network size (number of parameters), networks using Hironaka decompositions always outperform ones using minimal algebra generators which in turn outperform ones using the Weyl generators. 

The layout of the paper will be as follows. In Section \ref{sec:inv_systems}, we define the problem at hand. We follow this with an overview of the methods required to obtain generating sets of invariant polynomials in Section \ref{sec:generators}. In Section \ref{sec:UAT}, we discuss extensions of approximation theorems to the case of invariant functions to make use of these generators. In Section \ref{sec:testing}, we present a performance comparison of neural networks that utilise the invariant generators. We use nested sampling in Section \ref{sec:Bayesian} to compute model evidences and perform model comparison. Finally, we discuss the results and areas of possible future work in Section \ref{sec:discussion}.  

Similar work has been carried out in \cite{Lorentz-Top} but there, the authors rely on the finite-dimensional representations of the Lorentz group to construct Lorentz-equivariant networks and they deal with permutation invariance by imposing sums at the appropriate parts of the networks. 

\section{Setting the scene: invariant systems} \label{sec:inv_systems}

Let us begin by defining the problem more precisely. In data-driven approaches, one is tasked with describing a system which one has collected some data about. This data set is usually a collection of labelled pairs $D = \{ x_i, y_i \}$, where the $x_i \in X$ are a set of input variables which belong to some space $X$ and the $y_i \in Y$ are some known target variables which belong to some other space $Y$.\footnote{We restrict our discussion to the case of supervised learning for clarity, but these ideas are also applicable to the unsupervised case, where one lacks the knowledge of the property one is seeking. There, although we do not know what property the function $f$ will describe, we still know that it must be invariant under the symmetries of the system.} In particle physics, $X$ usually corresponds to the space of particle momenta and $Y$ corresponds to a quantity of interest such as a cross-section. It is then assumed that a property of the system is described by some ``true" function $f : X \rightarrow Y : x_i \mapsto y_i$ and the typical task of machine learning is to find a function $\hat{f} : X \rightarrow Y$ which approximates $f$ in some sense.

If we have no more information about the system at hand, then this completes our set-up and from here, one usually proceeds to pick an algorithm from the vast machine learning toolbox to apply it to the data. But, when working with systems that possess certain symmetries, we have additional {\em a priori} knowledge and there is more to be said about the function $f$. Consider a group $G$ acting on the space $X$ in some well-defined way, $G \times X \rightarrow X : (g,x) \mapsto g \circ x$ for $x \in X$ and $g \in G$. If a system is invariant under the action of $G$ then the functions $f$ describing its properties must also be $G$-invariant, $f(g \circ x) = f(x)$ for all $g \in G$. When attempting to use machine learning tools to approximate these properties, this $G$-invariance imposes a huge restriction on the possible candidates for $f$ (and hence on its approximators) and it would be extremely wasteful to ignore it. If we deprive our algorithms of this knowledge of invariance, a lot of time, effort, and (often scarce) data will be wasted rediscovering what we already know rather than discovering new insights.

One way to incorporate group invariance into our description relies on modifying the data input. This modification can be carried out in a couple of ways. The first, more brute-force, method is to create additional artificial data by acting with the symmetry group on the original data. The algorithm is then trained on the combined data set in the hope that it learns that the functions it is approximating are constant on group orbits. Apart from requiring additional, often significant, computational effort, this method has the added disadvantage that it is impossible to carry out completely when the symmetry group is continuous. Another method relies on reformatting the input data. Instead of na\"ively inputting the collected data, which might not be invariant under the group action, as it is into our models, we instead construct a new set of input variables which are invariant.\footnote{More precisely, we map $X$ to some $X^\prime$ by $\phi : X \rightarrow X^\prime : x \mapsto x^\prime$, where the action of $G$ on $X^\prime$ is trivial, {\em i.e.} $G \times X^\prime \rightarrow X^\prime : (g, x^\prime) \mapsto g \circ x^\prime = x^\prime$. We then define $\hat{f} : X^\prime \rightarrow Y$ and the approximator function becomes $\hat{f} \circ \phi: X \rightarrow Y$.} But, if this is done in any old arbitrary way (such as taking the \texttt{min$()$}, \texttt{max$()$}, or \texttt{sum$()$} of some combination of the inputs), then, even if the algorithm is non-linear and can deal with possibly missing information (which is not always guaranteed), it will still waste resources constructing the missing information from the arbitrary invariant inputs. So crucially, one should reformat the input data without incurring any information loss. This requires finding a set of invariant variables which are able to construct all possible group invariant functions that could describe the properties of our system.

Alas, the space of all invariant functions is too large to consider and we do not have a mathematical handle on it. That is, we do not know how to generate it. We thus restrict our attention to invariant polynomials which, when exploited via some approximation theorems, allow us to achieve the same goal. In particular, we highlight and prove some approximation theorems which show that invariant polynomials are indeed enough in the case of continuous functions invariant under compact groups and linearly reductive groups defined over the complex numbers. We also discuss the difficulties that arise when the groups are defined over the reals. 

But the characterisation of invariant polynomials, unlike that of all invariant functions, is more tractable and is made possible by the tools of invariant theory. In previous work \cite{LPIP}, we applied these tools in the setting of particle physics to develop systematic methods that furnish us with sets of generating variables for the Lorentz- and permutation-invariant polynomials in the particle momenta. In the following Section, we discuss the properties of these generators and give some explicit examples in cases of physical interest.

\section{Algebras of invariant polynomials} \label{sec:generators}

We will briefly review the concepts required to characterise invariant polynomials under certain groups. For a complete discussion and precise definitions, refer to \cite{LPIP}.

Let $V \cong \C^n$ be a complex\footnote{We complexify everything to make use of powerful theorems from invariant theory. Once the dust settles, one can, with some care, simply restrict to the real numbers since a generating set over the complex numbers is also a generating set over the reals.} $n$-dimensional vector space, with $x_i$ labelling its components in some basis. The set of all polynomials in these variables $x_i$ with complex coefficients forms an algebra over $\C$ which is usually denoted by $\C[x_1, \dots, x_n]$ or $\C[V]$ for short.\footnote{An algebra comes with the expected operations of addition (with an inverse) and multiplication (without an inverse) under which it is closed.} If a group $G$ acts on $V$ via some representation, $\rho: G \rightarrow GL(V)$, which is extended to the algebra of polynomials in the obvious way, then the set of invariant polynomials under the action of $G$ forms a subalgebra of $\C[V]$ which we denote by $\C[V]^G = \{f \in \C[V] \  |  \ g \circ f = f, \forall g \in G \}$. It turns out, due to some theorems from invariant theory \cite{dk}, that this invariant subalgebra is finitely generated if $G$ is a linearly reductive group. Most groups one encounters in physics, and indeed the Lorentz and permutation groups, are examples of such linearly reductive groups. 

In these cases, one may find sets of generating polynomials which come in two flavours. The first, called a set of {\em minimal algebra generators} (MAG), is a set of homogeneous invariant polynomials $ g_1, \dots, g_r \in \C[V]^G$ such that any $f \in \C[V]^G$ can be written as a polynomial in the MAG, $f = h(g_1, \dots, g_r)$. Minimality here is the condition that no generator may be expressed as a polynomial in the other remaining ones. These generators have a unique number and polynomial degrees but are {\em not} unique. Furthermore, they are in general {\em not} algebraically independent, the implications of which will be discussed in Section \ref{sec:UAT}.

The second is a called a {\em Hironaka decomposition} (HD). A HD of an algebra, $\C[V]^G = \bigoplus_i \eta_i \C[\theta_j]$, is a free-finitely generated module over a subalgebra $\C[\theta_j]$.\footnote{Finitely-generated implies that there are finitely many $\eta_i$'s and free indicates that the $\eta_i$'s form a linearly independent basis.} Without diving into the details, a HD of an invariant algebra results in a set of homogeneous invariant polynomials $\{\theta_j \}$ and $\{\eta_i \}$, termed the {\em primaries}\footnote{The primaries are algebraically independent.} and {\em secondaries}\footnote{The secondaries always contain the trivial secondary $\eta_1 = 1$.} respectively, such that any $f \in \C[V]^G$ can be written as linear sum of the secondaries with coefficients that are polynomials in the primaries, $f = \sum_i \eta_i h_i(\theta_j)$. Notably, although this decomposition is also {\em not} unique, any polynomial $f \in \C[V]^G$ is {\em uniquely} expressed in terms of a HD.

As with all good things, finding the generators of an invariant algebra in either flavour is in general hard. The methods developed in \cite{LPIP} to construct both the MAG and HDs rely on finding (partially in the case of MAG) a HD of some invariant algebra. 

\subsection{Lorentz- and permutation-invariant polynomials}

We now focus on the setting of particle physics. Here, our processes are collision events which are described by the momenta of particles. These are just elements of a (complex) $d$-dimensional vector space $V \cong \C^d$ and when working with systems of $n$ particles, the resulting polynomial algebra is $\C[V^n]$, where $V^n \cong \C^{nd}$. We are interested in the subalgebra which is invariant under the Lorentz group, which becomes $SO(d,\C)$ after complexification (or $O(d,\C)$ if one imposes parity invariance as well). Additionally, if some of the particles are identical, we are further interested in the subalgebras invariant under a subgroup of the permutation group $P \subseteq S_n$. The invariant algebras we wish to characterise are therefore $\C[V^n]^{SO(d) \times P}$ (or $\C[V^n]^{O(d) \times P}$). 

A well-known result due to Weyl \cite{Weyl} is that in the case of trivial permutation invariance (\text{i.e.} $\C[V^n]^{SO(d)}$), the MAG are formed by the set of all $n(n+1)/2$ dot products and $n$ choose $d$ contractions of momenta with the $d$-dimensional Levi-Civita epsilon tensor. These generators are not algebraically independent; there are relations between the epsilons and the dot products and, when $n > d$, there are even relations between the dot products themselves. The work done in \cite{LPIP} was first, to find HDs in the case of no permutation symmetry and second, to find MAG and HDs of the invariant algebras when a subgroup of the permutation group $P \subseteq S_n$ was included.

We will now give some examples of these MAG and HDs. We will denote the vectors (momenta) by $(p_1, \dots, p_n) \in V^n$ where the index labels the particles (not the components of the vector).

In the case of no permutation symmetry, when $n \leq d$, a HD of the invariant algebra $\C[V^n]^{SO(d)}$ is formed by taking the dot products as the set of primaries and the epsilons tensors (along with the trivial secondary $1$) as secondaries. But when $n > d$, the dot products are no longer algebraically independent and so cannot form a set of primaries. The solution to the first non-trivial case, $ n = d+1$, was found in \cite{LPIP}. In $d = 4$, this is when $n=5$, {\em i.e.} the invariant algebra is $\C[V^5]^{SO(4)}$ where $V^5 \cong \C^{4\times 5}$. Here, a HD can be formed by taking the primaries to be
\begin{align*}
& \theta^1_i =  p_1 \cdot p_1 + p_i \cdot p_i, \ \ 2 \leq i \leq 5, \\
& \theta^2_{ij} =  p_i \cdot p_j, \ \ 1 \leq i < j \leq 5, \\
\end{align*}
and the secondaries to be
\begin{align*}
& \eta^1_i =  (p_1 \cdot p_1)^i, \ \  0 \leq i \leq 4, \\
& \eta^2_{ijkl} = \epsilon(p_i, p_j, p_k, p_l), \ \ \forall i \neq j \neq k \neq l.
\end{align*} 

Things quickly become more complicated when permutation invariance is thrown into the mix. As an illustration, consider the case when one is again working in $d = 4$ dimensions but with $n = 4$ particles which come in identical pairs. Here, the symmetry group is a direct product $SO(4, \C) \times S_2 \times S_2$ and the invariant algebra is $\C[V^4]^{SO(4) \times S_2 \times S_2}$ where $V^4 \cong \C^{4\times 4}$. The MAG can be formed by a set of $21$ generators listed in Table \ref{table:MAGgens} of Appendix \ref{app:gens}. We can also obtain a HD by taking the $10$ primaries and $16$ secondaries listed in Table \ref{table:HDgens} of Appendix \ref{app:gens}.

We list the number of generators for each of the algebras tested in Sections \ref{sec:testing} and \ref{sec:Bayesian} in Table \ref{table:numgens}. An explicit list of the generators of invariant algebras of multiple physically interesting cases can be found on-line.\footnote{Available at: \url{https://github.com/WardHaddadin/Invariant-polynomials-and-machine-learning}.}

\def\arraystretch{2}%
\begin{table}[!htb]
\begin{center}
\begin{tabular}{ |c|c|c|c| } 
\hline
Invariant algebra & MAG & HD (primaries) & HD (secondaries)  \\ 
\hline
$\C[V^4]^{SO(4)}$ & $11$ & $10$ & $2$ \\
\hline
$\C[V^4]^{SO(4) \times S_2}$ & $16$ & $10$ & $8$ \\
\hline
$\C[V^4]^{SO(4) \times S_2 \times S_2}$ & $21$ & $10$ & $16$ \\
\hline
$\C[V^4]^{SO(4) \times S_3}$ & $33$ & $10$ & $72$ \\
\hline
$\C[V^5]^{SO(4)}$ & $20$ & $14$ & $10$ \\
\hline
\end{tabular}
\caption{\label{table:numgens} The number of minimal algebra generators and primaries and secondaries of Hironaka decompositions of invariant algebras tested in Sections \ref{sec:testing} and \ref{sec:Bayesian}.}
\end{center}
\end{table}

\section{Approximation theorems} \label{sec:UAT}

Armed with the generators of invariant algebras, we now move onto discussing how they can be exploited in machine learning via approximation theorems. We will work in the field of complex numbers $\C$, but some of these results are also valid when restricted to the reals $\R$.

The famous Stone-Weierstrauss (SW) theorem \cite{SW1, SW2}, which forms the cornerstone of approximation theory, states that any continuous function defined on a compact space can be approximated by a polynomial. The most general form of the SW theorem is very powerful.\footnote{The full theorem can be found in Appendix \ref{app:SW}.} Here, we state the special version concerned with complex-valued functions defined on compact subspaces of $\C^n$.
\begin{theorem}[Stone-Weierstrauss]
Let $X \subset \C^n$ be a compact subspace and denote by $C(X)$ the algebra of continuous complex-valued functions, $f : X \rightarrow \C$. The polynomial subalgebra $\C[X]$ is dense in $C(X)$. That is, for any $f \in C(X)$ and any real $\epsilon >0$, one can find a polynomial $r \in \C[X]$ such that $| r(x) - f(x) | < \epsilon$ for all $x \in X$, where $|-|$ is defined as the sup norm.
\end{theorem}  
The SW theorem allows us to approximate all continuous functions (and thus all continuous invariant ones) using the polynomials in the coordinates $x_i$ of $\C^n$ in some basis. We, however, would like to employ the MAG and HDs in our machine learning quests to approximate invariant functions. This crucially hinges on our ability to approximate all invariant functions using invariant polynomials which we can then express via MAG and HDs. Therefore, we must generalise (or rather restrict) the above theorem to the case of invariant functions and invariant polynomials. It turns out that this is non-trivial in general and we have not been able to demonstrate it completely. We sketch the roadmap to a full proof and discuss the bottlenecks that arise in Appendix \ref{app:SW}.

The generalisation for {\em analytic} invariant functions is trivial since they all admit power series representations. Inside their radii of convergence, these power series can be truncated into polynomials, which are furthermore invariant, to obtain arbitrarily accurate approximations of the analytic functions. But when working with {\em continuous} invariant functions, we do not have the luxury of power series representations and so we must resort to other means. In the case of continuous real-valued functions invariant under compact groups, the problem is relatively straightforward and has been solved in \cite{Yarotsky}. Here, we will prove the generalisation for the case of continuous complex-valued functions invariant under linearly reductive groups defined over $\C$. To do this, we first state a result which can be found in \cite{dk} or \cite{Kraft}.
\begin{theorem} \label{thm:reductive}
Let $G$ be a linearly reductive group defined over $\C$ and let $H$ be a maximal compact subgroup $H \subset G$. Consider the algebras of invariant polynomials $\C[V]^H$ and $\C[V]^G$. Then, $\C[V]^H = \C[V]^G$. 
\end{theorem}
The above theorem does not hold for linearly reductive groups defined over $\R$.\footnote{This is because the result relies crucially on the fact that when $G$ is a linearly reductive group defined over $\C$, the maximal compact subgroup is Zariski-dense in $G$. This is not guaranteed when $G$ is defined over $\R$.} The following arguments therefore only restrict to linearly reductive groups $G$ defined over $\R$ if $G$ itself is compact.

We also require the concept of a (normalised) Haar measure of a compact group $H$, $\mu_H$, which allows us to define an integral over the group, $\int_{h \in H} d\mu_H$. By borrowing the arguments from \cite{Yarotsky}, we can now prove our generalisation as follows.
\begin{theorem} \label{thm:inv_poly}
Let $G$ be a linearly reductive group over $\C$ acting on a finite-dimensional vector space $V \cong \C^n$ and denote by $\C[V]^G$ the algebra over $\C$ of $G$-invariant polynomials. Then, any continuous complex-valued $G$-invariant function $f : V \to \C $ can be approximated by an invariant polynomial $r_{sym} \in \C[V]^G$ on a compact subspace $X \subset V$.
\end{theorem}

{\em Proof:} 
Let $X$ be a compact subspace of $V$ and consider the maximal compact subgroup $H \subset G$. Further consider the symmetrized set $X_{sym} = \cup_{h \in H} (h \circ X)$ which is compact as it is the image of a compact set $H \times X$ under a continuous map. By the Stone-Weierstrass theorem, for any $\epsilon > 0$ there exists a polynomial $r \in \C[V]$ such that $|r(x) - f(x)| < \epsilon$ for $x \in X_{sym}$. But $r$ might not be $G$-invariant. So instead, consider the symmetrized function $r_{sym}(x) = \int_{h \in H} r(h \circ x) d\mu_H$ (where $h \circ x \in X_{sym}, \forall h \in H$)  which is $H$-invariant. We note that $r_{sym}$ is a polynomial, since $r(h \circ x)$ is a fixed degree polynomial in $x$ for any $h \in H$, and therefore $r_{sym} \in \C[V]^H$. By Theorem \ref{thm:reductive}, we consequently have that $r_{sym} \in \C[V]^G$. Now, 
\begin{align*}
|r_{sym}(x) - f(x)| & = \left| \int_{h \in H} r(h \circ x) - f(h \circ x) d\mu_H \right| \\
& \leq \int_{h \in H} |r(h \circ x) - f(h \circ x)| d\mu_H < \int_{h \in H}  \epsilon d\mu_H  = \epsilon
\end{align*}
for all $x \in X$, where we used the fact that $f$ is $G$-invariant (and thus $H$-invariant) and that $|r(x) - f(x)| < \epsilon$. $\hfill \Box$

Using this theorem, we can now use the generators of invariant polynomials (MAG and HDs) in machine learning algorithms to approximate continuous invariant functions. For the rest of the discussion, we will restrict our attention to the infamous machine learning tool, neural networks. It is a well-known fact that a neural network with a (large enough) single layer and a non-polynomial activation function can approximate any continuous real-valued function defined on some compact subspace of $\R^n$ \cite{Pinkus}. Recently, a generalisation to complex-valued functions was proved in \cite{Voigtlaender}.

\begin{theorem}[Voigtlaender] \label{thm:Voigtlaender}
Let $\sigma : \mathbb{\C} \to \mathbb{\C}$ be a function that is {\em not} almost polyharmonic.\footnote{A function $\sigma$ is almost polyharmonic if there exist $m \in \mathbb{N}$ and an infinitely differentiable $g: \C \rightarrow \C$ with $\Delta^m g = 0$ such that $\sigma = g$ almost everywhere. Here, $\Delta = \frac{\partial^2}{\partial x^2} + \frac{\partial^2}{\partial y^2}$ is the usual Laplace operator on $\C \cong \R^2$.}  Let $V \cong \mathbb{C}^n$ be a complex finite-dimensional vector space. Then any continuous function $f : V \to \mathbb{\C}$ can be approximated, in the sense of uniform convergence on compact sets\footnote{That is, for any compact subspace $X \subset V$ and any real $\epsilon >0$, one can find a function $\hat{f}$ such that $| f(x) - \hat{f}(x) | < \epsilon$ for all $x \in X$. Here, the norm $|-|$ is defined by the uniform, or sup, norm.}, by functions $\hat{f}: V \to \mathbb{\C}$ of the form 
\begin{align*}
\hat{f}(x) = \sum_{i=1}^N c_i \sigma \left( \sum_{s = 1}^n w_{is} x_s + h_i \right)
\end{align*}
for $x \in V$, some parameter $N$, and complex coefficients $c_i, w_{is}, h_i$.
\end{theorem}
The above theorem is for a complex-valued neural network with a single layer of width $N$. Interestingly, unlike real-valued neural networks, the criterion on the activation function for single layer (shallow) and multiple layer (deep) complex-valued networks is different.\footnote{For multiple layer neural networks, $\sigma$ must be neither a polynomial, a holomorphic function, or an antiholomorphic function.}\footnote{This theorem can also be used to approximate any continuous complex-valued multi-dimensional function $f:V \rightarrow \C^m$ by ``stitching" together $m$ networks each approximating one of the components of $f$.}

In \cite{Yarotsky}, real-valued neural networks were shown to be able to approximate any continuous real-valued function invariant under a compact group using the MAG of the invariant algebra of polynomials as inputs. Here, we will prove the generalisation of that theorem to complex-valued neural networks and continuous complex-valued functions invariant under linearly reductive groups defined over $\C$.

\begin{theorem}[MAG networks] \label{thm:MAG}
Let $G$ be a linearly reductive group over $\C$ acting on a finite-dimensional vector space $V \cong \C^n$ and let $g_1,\dots, g_m : V \to \C$ be a set of minimal algebra generators of the invariant algebra $\C[V]^G$. Then, any continuous $G$-invariant function $f : V \to \mathbb{\C}$ can be approximated, in the sense of uniform convergence on compact sets, by invariant functions $\hat{f}: V \to \mathbb{\C}$ of the form
\begin{align*}
\hat{f}(x) = \sum_{i=1}^N c_i \sigma \left( \sum_{s = 1}^m w_{is} g_s(x) + h_i \right)
\end{align*}
for $x \in V$, some parameter $N$, and complex coefficients $c_i, w_{is}, h_i$.
\end{theorem}

{\em Proof:} It is obvious that the functions $\hat{f}$ are $G$-invariant, so we only need to prove completeness.

From theorem \ref{thm:inv_poly}, we can approximate $f$ by an invariant polynomial $r_{sym} \in \C[V]^G$ on a compact subspace $X \subset V$. Next, we use the MAG to express $r_{sym} = h(g_1, \dots, g_m)$ for some polynomial $h$. It remains to approximate the polynomial $h(g_1, \dots, g_m)$ by an expression of the form $H(x) = \sum_{i=1}^N c_{i} \sigma( \sum_{s = 1}^m w_{is} g_s (x) + h_i)$ on the compact set $\{g_1(x), \dots , g_m(x) | x \in X\}$. Using Theorem \ref{thm:Voigtlaender}, we can do this with accuracy $\epsilon^\prime$, $|H(x) - h(x)| < \epsilon^\prime$ for all $x \in X$. Finally, setting $\hat{f}(x) = H(x)$, we obtain an $\hat{f}$ of the required form such that $|\hat{f}(x) - f(x)| < \epsilon$ for $x \in X$. $\hfill \Box$

As we have mentioned in Section \ref{sec:generators}, the MAG are not algebraically independent in general. This leads to redundancies in the description of invariant polynomials via the MAG which, consequently, reduces the efficiency of neural networks using the MAG as inputs. HDs however are redundancy-free and express any invariant polynomial uniquely. To make use of this, we generalise Theorem \ref{thm:MAG} to the case of HDs as follows.
\begin{theorem}[HD networks] \label{thm:HDNET}
Let $G$ be a linearly reductive group over $\C$ acting on a finite-dimensional vector space $V \cong \C^n$ and let $\theta_1, \dots , \theta_m : V \to \C$ and $\eta_1, \dots , \eta_p : V \to \C$ be a Hironaka decomposition (primaries and secondaries respectively) of the invariant algebra $\C[V]^G = \bigoplus_i \eta_i \C[\theta_j]$. Then, any continuous $G$-invariant function $f : V \to \C$ can be approximated, in the sense of uniform convergence on compact sets, by invariant functions $\hat{f}: V \to \C$ of the form
\begin{align*}
\hat{f}(x) = \sum_{k=1}^p \eta_k(x) \sum_{i=1}^N c_{ki} \sigma \left( \sum_{s = 1}^m w_{is} \theta_s (x) + h_i \right)
\end{align*}
for $x \in V$, some parameter $N$, and complex coefficients $c_{ki}, w_{is}, h_i$.
\end{theorem}

{\em Proof:} It is obvious that the functions $\hat{f}$ are $G$-invariant, so we only need to prove completeness.

From theorem \ref{thm:inv_poly}, we can approximate $f$ by an invariant polynomial $r_{sym} \in \C[V]^G$ on a compact subspace $X \subset V$. Next, we use a HD to express $r_{sym} = \sum_{k = 1}^p \eta_k h_k(\theta_1, \dots, \theta_m)$ uniquely. It remains to approximate the polynomials $h_k(\theta_1, \dots, \theta_m)$ by expressions of the form $H_k(x) = \sum_{i=1}^N c_{ki} \sigma( \sum_{s = 1}^m w_{is} \theta_s (x) + h_i)$ on the compact set $\{\theta_1(x), \dots , \theta_m(x) | x \in X \}$. Using the multi-dimensional version of Theorem \ref{thm:Voigtlaender}, we can do this with accuracy $\epsilon^\prime$, $|H_k(x) - h_k(x)| < \epsilon^\prime$ for all $k$ and $x \in X$. Finally, setting $\hat{f}(x) = \sum_{k=1}^p \eta_k(x) H_k(x)$, we obtain an $\hat{f}$ of the required form such that $|\hat{f}(x) - f(x)| < \epsilon$ for $x \in X$. $\hfill \Box$

As we will see in Section \ref{sec:Bayesian}, the width of the network, $N$, is very important in HD networks. It must be large enough and comparable to the number of secondaries so that the network can approximate the (in general distinct) polynomials $h_k(\theta_1, \dots, \theta_m)$. When this is satisfied, the full power of HD networks becomes apparent over MAG networks.

In the following Section, we compare the performance of the networks in Theorems \ref{thm:Voigtlaender}, \ref{thm:MAG}, and \ref{thm:HDNET} under a range of hyperparameters. A basic \texttt{C++} implementation of these networks can be found on-line.\footnote{Available at: \url{https://github.com/WardHaddadin/Invariant-polynomials-and-machine-learning}.}

\section{Brute-force testing} \label{sec:testing}

We will perform our testing by applying the above networks to regression problems.\footnote{We restrict to real-valued functions throughout our testing.} That is, given some data set $D = \{ x_i, y_i \}$ described by a true function $f$, we would like to train our networks, $\hat{f}$, by tuning their coefficients to approximate $f$. Goodness-of-fit will be defined here using the loss metric of {\em mean squared error}, $\text{MSE}(D) = \frac{1}{|D|} \sum_{i} (\hat{f}(x_i) - y_i)^2$. Instead of performing regression on arbitrary invariant functions, it is a better idea to do regression on polynomials.\footnote{There is some theoretical work on networks fitting polynomials \cite{Poly-nets1, Poly-nets2} which discusses and proves some results regarding the approximating power of networks in relation to polynomial degree and sparsity.} This is because by regressing on polynomials of a certain maximum degree, we are able to control the ``difficulty" of the regression problem which in turn allows us to quantify the performance of the networks when applied to problems of varying complexity. Additionally, we can easily produce randomly generated polynomials for testing. 

We consider the Lorentz- and permutation-invariant polynomials in particle momenta and perform testing for the invariant algebras listed in Table \ref{table:numgens}. It is obvious that using the dot products and epsilon tensor contractions as inputs is a better idea than just using the plain momenta.\footnote{Indeed, as far as we are aware, no computer has yet discovered Lorentz invariance by itself. But, given an arbitrary symmetric metric $g_{\mu \nu}$, a neural network with inputs $g_{\mu\nu}p_i^\mu p_j^\nu$ can be trained to converge on the Minkowski metric \cite{NNmetric}.} We therefore restrict our testing to compare the performance of networks utilising HDs and MAG, which we call {\em HD} and {\em MAG networks} respectively, against networks utilising the Weyl generators (plain dot products and epsilon tensor contractions), or {\em Weyl networks}.

\subsection{Summary of results}

For each hyperparameter, testing was carried out on $100$ repetitions of randomly generated invariant polynomials and randomly generated data.\footnote{The input and output data was normalised to be in the range $[-1,1]$.} The data was split into an $80:20$ train-to-validation ratio and $L_2$-regularisation methods were used to prevent overfitting (this is discussed in the context of Bayesian neural networks in Section \ref{sec:Bayesian}).

\begin{sidewaysfigure}[!]
	\centering
    \includegraphics[width=\textwidth]{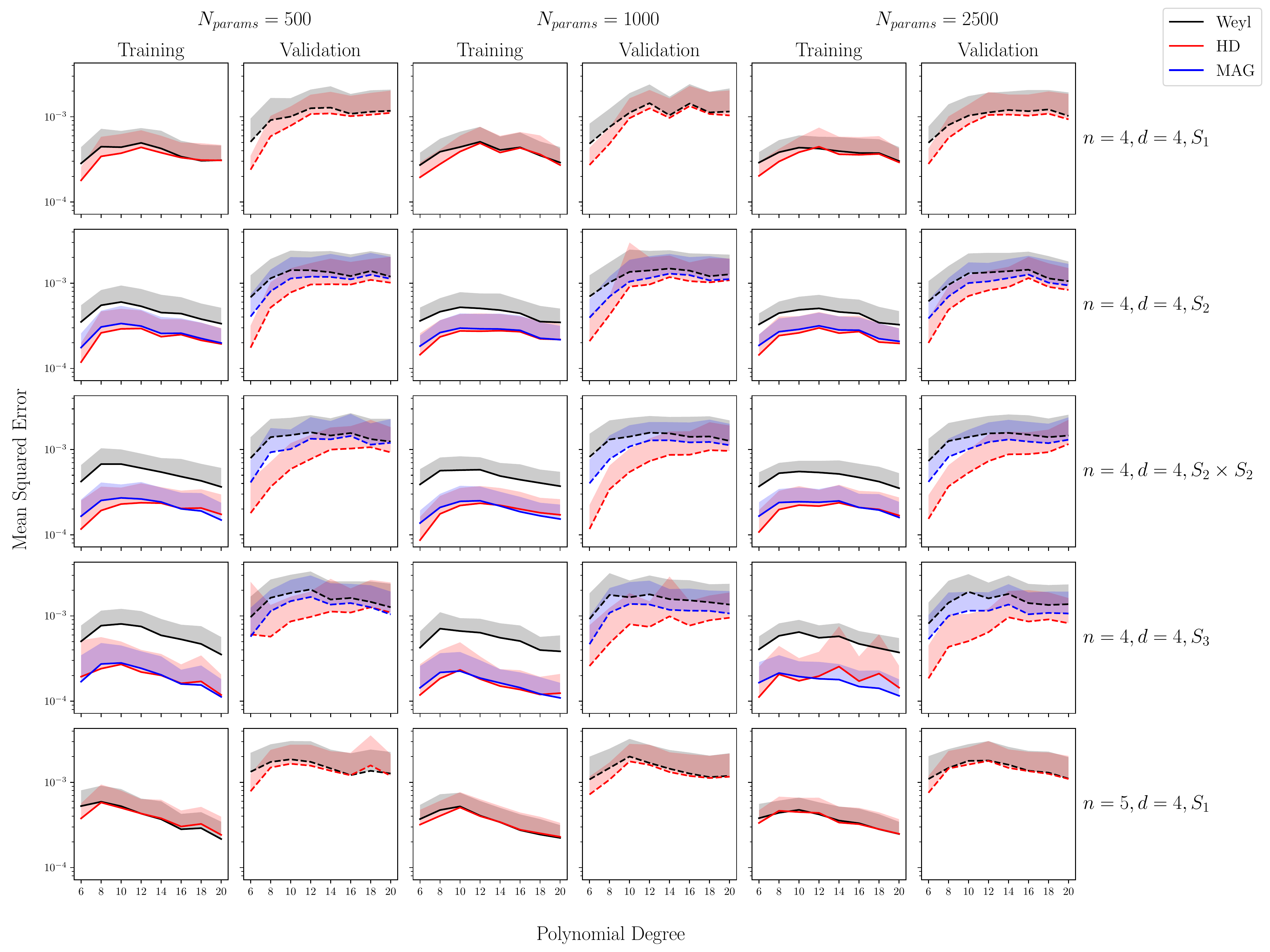}
    \caption{A grid showing the performance of Hironaka decomposition, minimal algebra generators, and Weyl networks applied to polynomial regression with increasing polynomial degree. Averaged from $100$ runs on randomly generated polynomials at each degree. $L_2$ regularisation was applied with $\lambda = 10^{-5}$. The activation function of the hidden layers was tanh$(x)$ and the number of hidden layers was $2$.}\label{fig:deg}
\end{sidewaysfigure}

\begin{sidewaysfigure}[!]
	\centering
    \includegraphics[width=\textwidth]{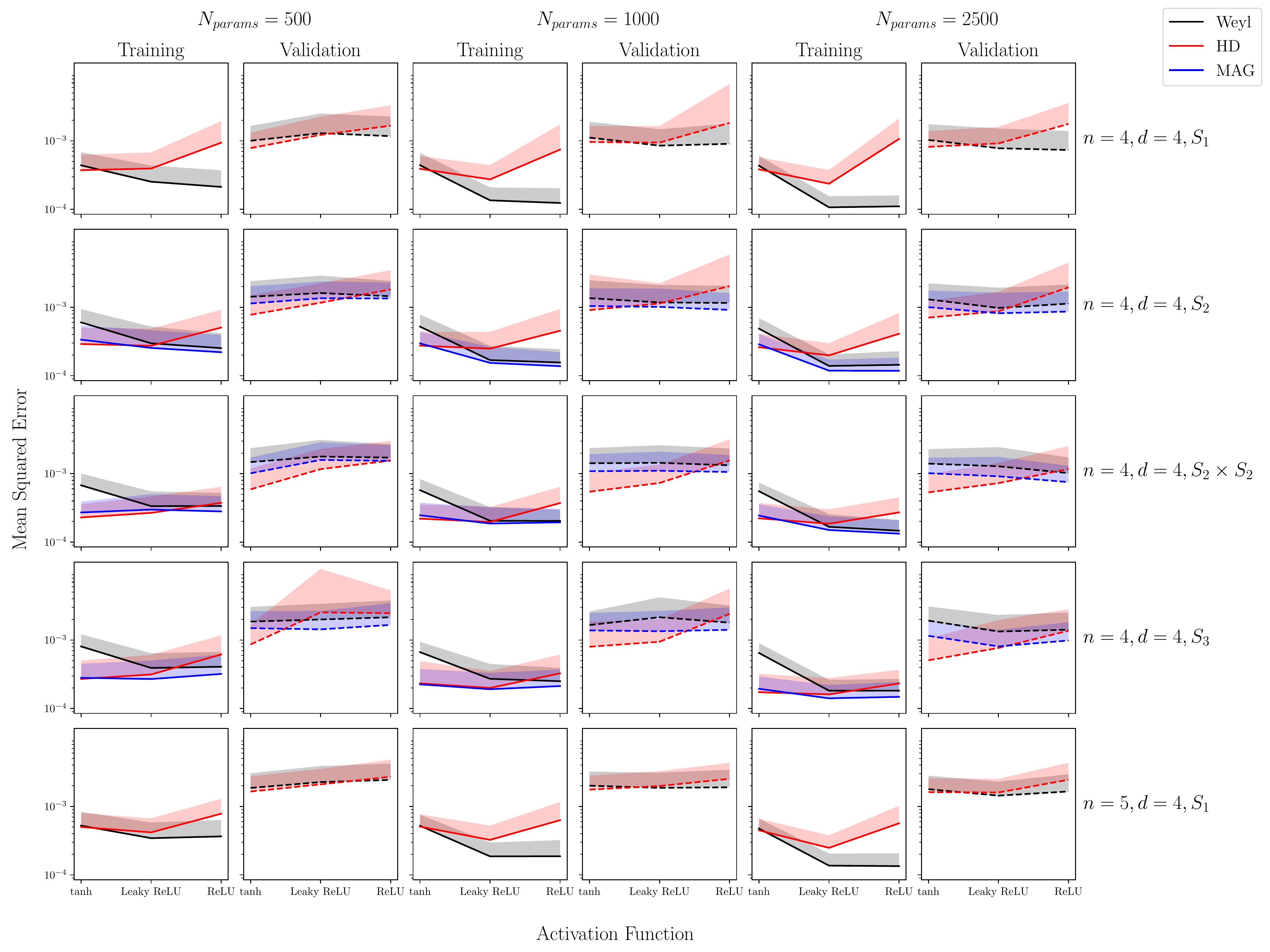}
    \caption{A grid showing the performance of Hironaka decomposition, minimal algebra generators, and Weyl networks for some invariant algebras applied to polynomial regression with different activation functions. Averaged from applying the networks to $100$ randomly generated polynomials for each activation function. $L_2$ regularisation was applied with $\lambda = 10^{-5}$. The polynomial degree used was $10$ and the number of hidden layers was $2$.}\label{fig:act}
\end{sidewaysfigure}

\begin{sidewaysfigure}[!]
	\centering
    \includegraphics[width=\textwidth]{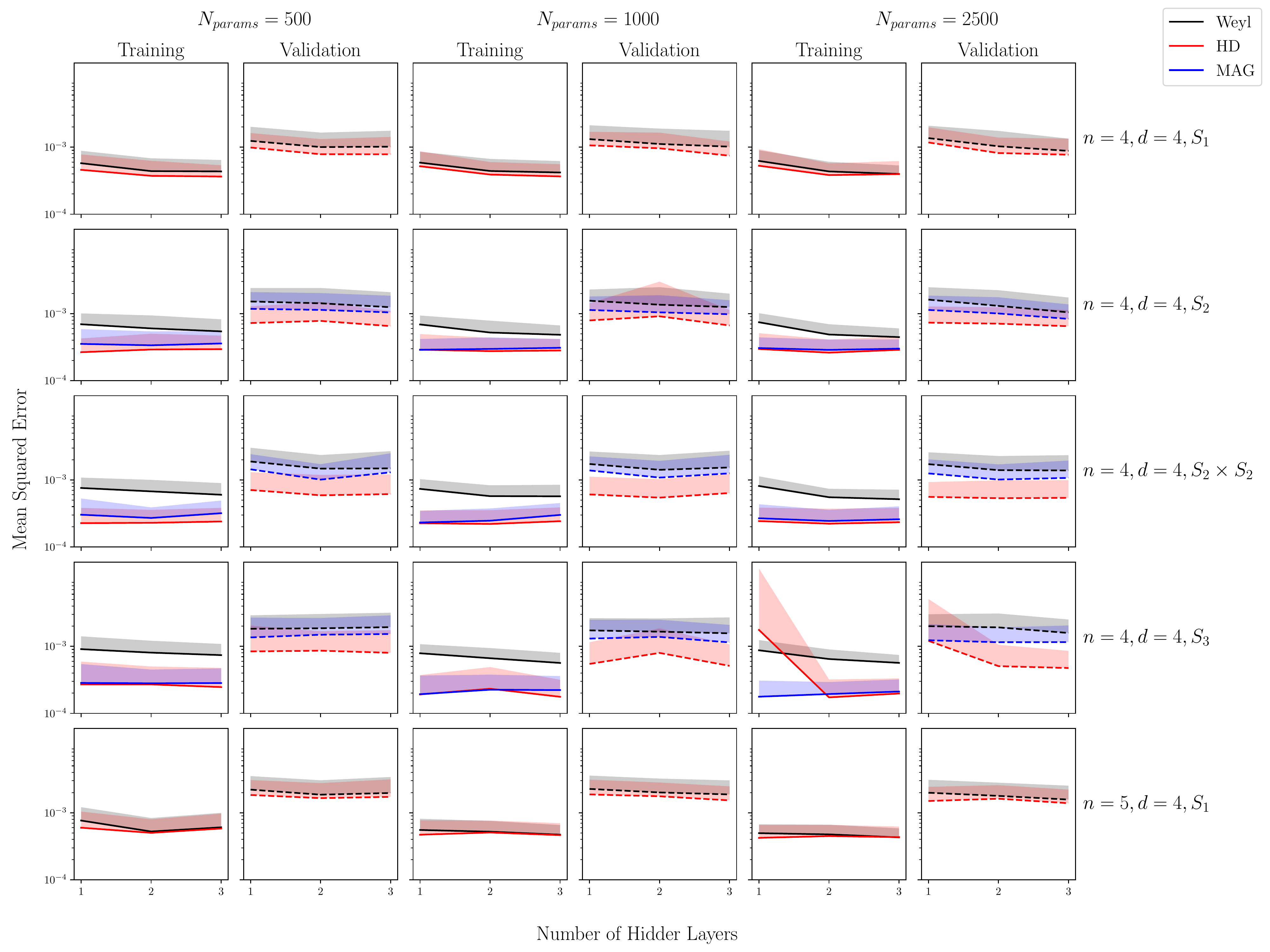}
    \caption{A grid showing the performance of Hironaka decomposition, minimal algebra generators, and Weyl networks for some invariant algebras applied to polynomial regression with increasing number of hidden layers. Averaged from applying the networks to $100$ randomly generated polynomials for each activation function. $L_2$ regularisation was applied with $\lambda = 10^{-5}$. The activation function of the hidden layers was tanh$(x)$ and the polynomial degree used was $10$.}\label{fig:layers}
\end{sidewaysfigure}

The general theme that can be seen from the results in Figures \ref{fig:deg}, \ref{fig:act}, and \ref{fig:layers} is that HD networks perform better that MAG networks which in turn perform better that Weyl networks. The improvement in performance becomes more apparent in the presence of permutation groups of higher order. We can also see that the improvement is more significant with lower degree polynomials, and while still present at higher degrees, plateaus. Notably, HD networks tend to prefer tanh activation functions while MAG and Weyl networks perform better with ReLU and Leaky ReLU. 

Below, we provide a more detailed description of the behaviour of the networks under a variety of hyperparameters.


\subsubsection{Number of network parameters}

We tested how the benefit or dis-benefit of each network scales with the total number of parameters. That is, at what size does the investment in the MAG and HD networks become worth it? We will get a more complete picture of this in the next Section, where we treat these networks as Bayesian models. Here, we tested networks with $500$, $1000$, and $2500$ parameters.\footnote{The actual number of parameters used in testing does not always exactly match these values. The exact number depends on the number of inputs, the number of layers, and the width of the layers. We constructed the networks to get the actual number of parameters as close as possible to these values.}

We can see that there is no notable improvement in performance for MAG over Weyl networks as the number of parameters increases. But for HD networks, the efficiency gain increases with the number of parameters. This is probably due to the point made in Section \ref{sec:UAT}, that HD networks need to have a width comparable to the number of secondaries for their full power to kick in. 

\subsubsection{Activation functions}

Activation functions play a major role in the performance of networks. We tested some of the most popular activation functions
\begin{align*}
& \text{tanh}(x) \\
& \text{ReLU}(x) = \text{max}(x,0) \\
& \text{Leaky ReLU}(x) = \text{max}(x,0) + \text{max}(0,-c x)
\end{align*}
where $c$ is some small positive constant. Since we are doing a regression task, the activation function of the output layer will always be a linear activation function. What we vary are the activation functions of the hidden layers. 

A common problem which arises with the ReLU function is called ``ReLU death". This occurs when the network parameters get pushed by gradient descent to the region where the ReLU function is zero and so cannot leave (since the gradient there is also zero). The Leaky ReLU tries to combat this issue by adding a small negative contribution in that region.

We can see that in all cases in Figure \ref{fig:act}, HD networks prefer tanh$(x)$ as an activation function. We can also see that the HD networks perform poorly with ReLU activation functions. More testing is required to understand why this happens exactly, but it might be due to the final secondaries layer being more susceptible to ReLU death.

\subsubsection{Depth vs. Width of networks}

We also tested how the depth of the networks affects performance. Generically, the touted mantra is that shallower networks (with fewer hidden layers) learn faster but deeper networks (with more hidden layers) learn more complicated structures in the data. To see if this applies here, we tested networks with $1, 2$, and $3$ hidden layers. 

From Figure \ref{fig:layers}, we can see that the width vs. depth ratio makes little difference. For HD networks, this is true as long as the width of the networks is comparable to the number of secondaries. As we will see in Section \ref{sec:Bayesian}, if given the choice between narrow and deep vs. wide and shallow networks, one should always choose the wide and shallow ones to allow the networks to better approximate the polynomials $h_k(x)$.

\subsubsection{Size of data set}

Another interesting hyperparameter to test was the size of the data set. That is, how much data does each type of network need to learn? This is very important when one is trying to probe properties of the system which either occur rarely (and so appear less frequently in the data) or when one does not have much data about the system because it is expensive or difficult to obtain. We carried out the testing using $5000$, $7000$, and $10,000$ data points. As expected, training on a larger data set resulted in lower overall training loss and validation loss. But, we found no additional significant improvement in any one network over the others.

\subsubsection{Error of data set}

Finally, it was also interesting to see how noise in the data set affects performance. That is, are some architectures more robust to noise than others? We trained the networks on data sets with $0\%$ error ({\em i.e.} ``perfect" data), $1\%$ error, $5\%$ error, and for an extreme case $20\%$ error. As expected, the noisier data sets resulted in worse performance, but, as with the size of the data set, there was no notable reduction of the performance of one network over the others.

\section{A Bayesian's-eye view} \label{sec:Bayesian}

\begin{figure}[!]
	\centering
    \includegraphics[width=0.6 \textwidth]{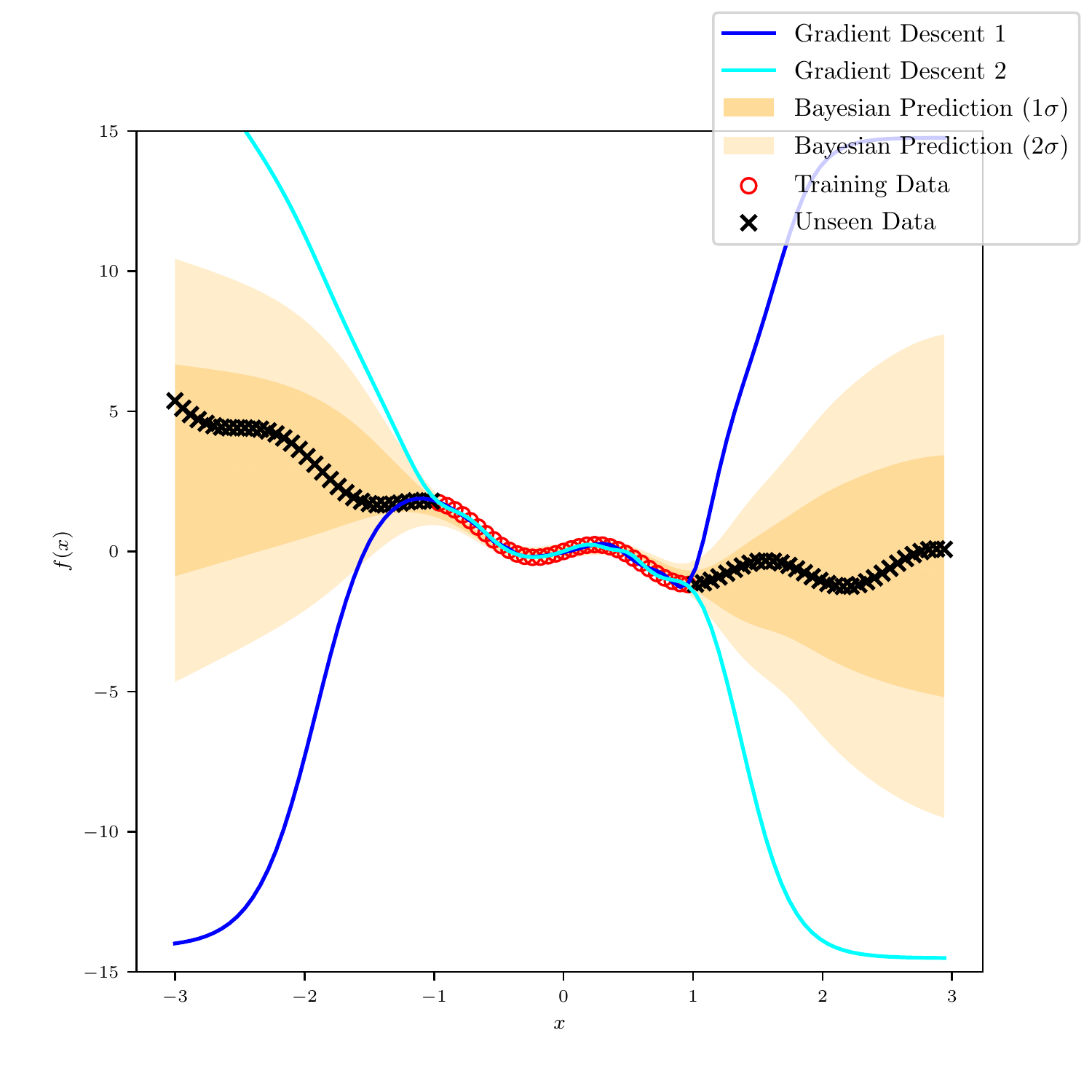}
    \caption{Neural networks applied to a regression problem. The predictions from two networks trained using the conventional approaches of gradient descent fit the training data well but perform poorly on unseen data. The Bayesian approach on the other hand provides meaningful and modest predictions on unseen data, while still performing well on training data.}\label{fig:evidence_example}
\end{figure}

Let us now take a step back and re-examine what comparing neural networks actually means. A neural network is essentially a glorified model, which we denote by $M$, that depends on some parameters $\alpha = \{ \alpha_i \}$. ``Training" a neural network amounts to tuning these parameters, using some gradient descent algorithm, to fit the data ``best", where goodness-of-fit is judged by some loss function. More precisely, this process can be viewed as the {\em maximum likelihood estimation} (MLE) of a Bayesian inference problem on some data set $D$. Let $P(D|\alpha, M)$ denote the likelihood\footnote{The likelihood must be a positive semi-definite, strictly decreasing, function of the loss.} of the data given some parameters $\alpha$ and let $P(\alpha|M)$ denote the prior\footnote{The prior corresponds to using {\em regularisation} in conventional training approaches.  A gaussian prior on the parameters corresponds to using $L_2$-regularisation.} on these parameters. Then, what is usually termed ``training" or ``learning" is the process of finding the MLE of the posterior
\begin{align} \label{eq:posterior}
P(\alpha | D,M) \propto P(D|\alpha, M) P(\alpha|M).
\end{align}
A glaring disadvantage of thinking solely of the MLE as the end goal is the fact that we gain no insight of how confident our model (neural network) is performing when applied to unseen data. Two different points in parameter space might both result in a high posterior ({\em i.e.} perform well on training data), but correspond to models whose performance is very different on unseen data. For a toy example, consider Figure \ref{fig:evidence_example}. Although the network fits the training data very well for the two points in parameter space, the behaviour on unseen data is very different. Furthermore, both of these choices of parameters perform poorly on unseen data.

Therefore, instead of thinking of the MLE as the ultimate goal, a model scientist\footnote{Pardon the pun.} should think of neural networks as Bayesian models. Viewed as such, the predictions of Bayesian neural networks, obtained by averaging the predictions of a regular network at all points of the parameter space weighted by the posterior (corresponding to goodness-of-fit) as
\begin{align} \label{eq:int}
\hat{f}_\text{avg}(x) = \int \hat{f}(x| \alpha) P(\alpha | D,M) d\alpha,
\end{align}
where $\hat{f}(x| \alpha)$ is the output of the network for some set of parameters $\alpha$, are now much more meaningful.\footnote{One can also quantify the uncertainty here as well by computing the variance Var$(\hat{f}(x)) = \int \hat{f}(x| \alpha)^2 P(\alpha | D,M) d\alpha - \hat{f}_\text{avg}(x)^2$.} The advantage of this procedure can be seen in Figure \ref{fig:evidence_example}, where the Bayesian network gives better predictions, but more importantly is modest where it is uncertain ({\em i.e.} on the unseen data). Alas, this approach is not without its limitations. For a thorough discussion of the issues that arise, see \cite{HandleyBNN}. Here, we will highlight two immediate problems that come to mind. 

The first, and major, problem is twofold and can be traced back to the ever-looming curse of dimensionality. When working with models with more than a few hundred parameters, the integral in Equation \ref{eq:int} quickly becomes intractable. On one hand, high dimensional integrals are difficult to perform, but also, computing the required constant of proportionality of Equation \ref{eq:posterior}
\begin{align*}
Z(M) = P(M | D) = \int P(D|\alpha, M) P(\alpha|M) d\alpha,
\end{align*} 
termed the {\em evidence}, is equally as difficult. The usual way around both of these problems is to use {\em Markov chain Monte Carlo} (MCMC) methods which produce a representative sample of points from the posterior and avoid computing $Z(M)$. The integral in Equation \ref{eq:int} then reduces to a manageable sum and one can obtain an estimate of the function $\hat{f}_\text{avg}(x)$. But since these methods have no access to $Z(M)$, they miss out on the opportunity to compare different models (in our case networks utilising different input methods). This is because, given the evidences, $Z(M_i)$, and priors, $P(M_i)$, of a set of models $\{ M_i \}$, the posterior (or probability) of a model is given by
\begin{align*}
P(M_i| D) =  \frac{P(D|M_i) P(M_i)}{P(D)} = \frac{Z(M_i) P(M_i)}{\sum_i Z(M_i) P(M_i)}.
\end{align*} 
Therefore, having access to the evidence of models gives us an excellent tool to choose, with good confidence, the best performing model (or even better, average the predictions from all the models weighted by their respective posteriors). One of the cutting-edge methods used to compute the evidence is called {\em nested sampling}. Its basic premise is to transform the multi-dimensional integral in Equation \ref{eq:int} to a one-dimensional one which is much easier to compute. We give a brief exposition of nested sampling in Appendix \ref{app:NS}.

The second obstacle that we face when applying this Bayesian approach to neural networks is that unlike usual models where one knows what the likelihood is, here we have the freedom to choose the likelihood. This extra freedom seems to throw us back to square one, where we are forced to choose a preferred likelihood function over any other function.\footnote{There have been some recent atempts to perform likelihood-free inference \cite{Like-free} but we will not discuss these here.} In the case of a mean-square-error loss function, this is slightly remedied by the fact that we can make an educated choice and construct a gaussian looking function of the loss
\begin{align*}
P(D|\alpha, M) \propto \text{exp}\left( - \frac{\sum_i (y_i - \hat{f}(x_i))^2}{2 N \sigma^2} \right),
\end{align*}
to act as the likelihood. There is still the freedom of choosing the parameter $\sigma^2$ though, but since we are only concerned with performing this analysis as a proof-of-concept, we will make the simplifying choice of setting $\sigma^2 = 1$.\footnote{The Bayesian way to deal with this issue is to make $\sigma^2$ itself a parameter of the model which one then infers from the data. The process of inferring hyperparameters from the data is called {\em Hierarchical modelling}.}

We will brush this second problem under the rug for the rest of this work and focus on obtaining evidence estimates for our networks using the semi-arbitrary gaussian choice of the likelihood. To perform this analysis, we used the nested sampling package \textsc{PolyChord} \cite{Polychord1, Polychord2}.\footnote{Available as \textsc{PolyChordLite} on \texttt{github}: \url{https://github.com/PolyChord/PolyChordLite}} In line with the usual $L_2$-regularisation techniques used in gradient descent algorithms, we choose a gaussian prior on the parameters 
\begin{align*}
P(\alpha| M) \propto \text{exp}\left( - \frac{\sum_i \alpha_i^2}{2 \sigma_{\alpha}^2} \right)
\end{align*}
where $\sigma_{\alpha}^2 = 0.01$. Unfortunately, the curse of dimensionality, although mitigated by nested sampling, still haunts us. We therefore only perform this analysis on networks of sizes $100$, $200$, and $300$ parameters with a single hidden layer.

\begin{figure}[!]
	\centering
    \includegraphics[width=\textwidth]{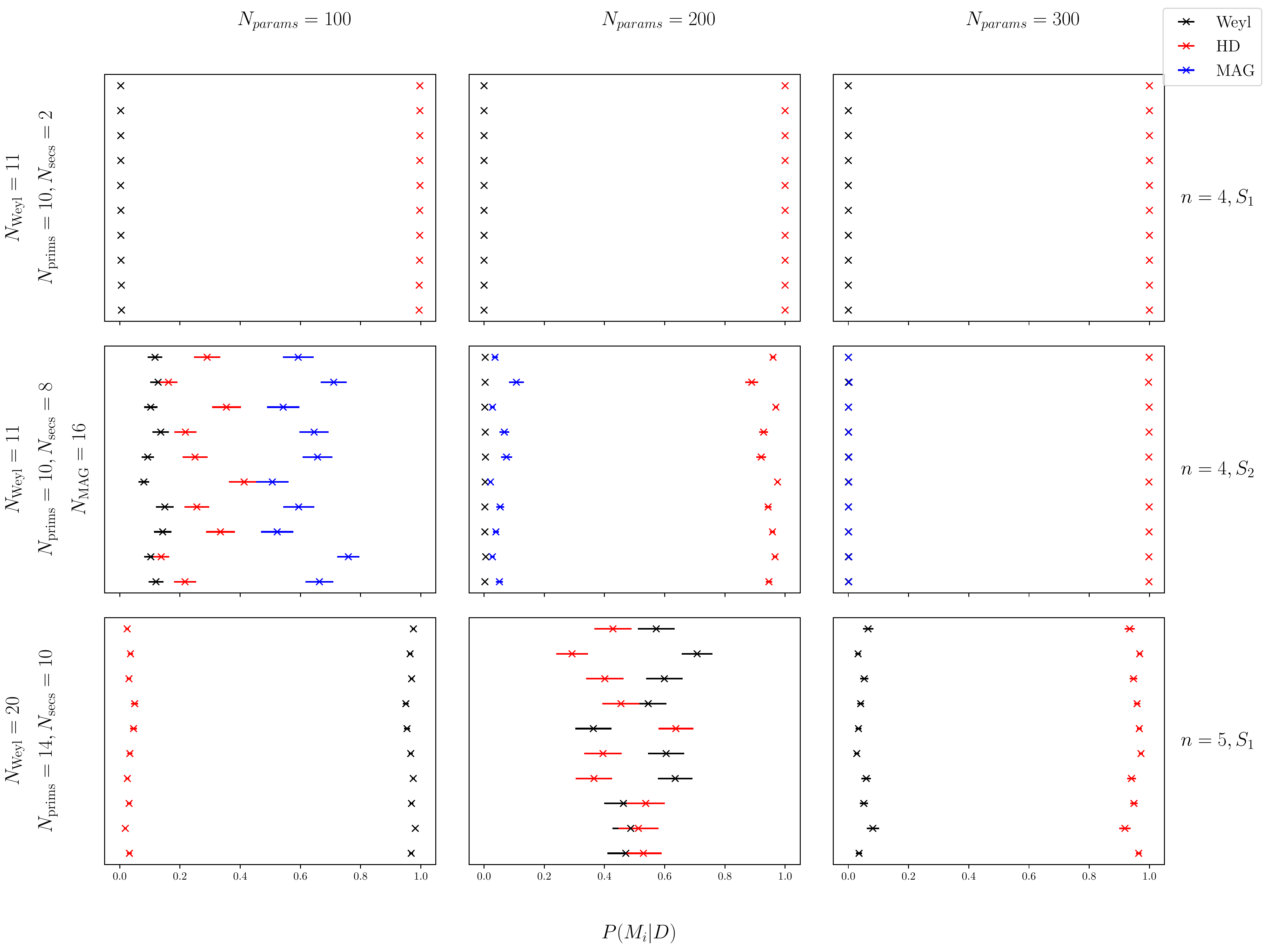}
    \caption{A grid showing the evidence of Hironaka decomposition, minimal algebra generators, and Weyl networks with a single layer and an increasing number of generators applied to polynomial regression. The evidence of each network was computed $10$ times. The activation function of the hidden layer was tanh$(x)$ and the polynomial degree used was $10$.}\label{fig:evidences}
\end{figure}

From Figure \ref{fig:evidences}, we can see that the performance of the HD networks depends highly on the number of parameters. As discussed previously, and further confirmed here, this is because for the HD networks to make full use of the nice structure of a HD, they must have enough nodes ($\sim O(\text{number of secondaries})$) in the networks to approximate the polynomials in the primaries multiplying the secondary generators. After a certain threshold, their performance significantly overtakes that of MAG networks which in turn perform better than Weyl networks. These results, combined with those found in Section \ref{sec:testing}, seem to confirm what was found in \cite{Mackay}, namely that networks with a higher evidence tend to perform better on unseen validation data in general. 

Although their evidence and validation loss does not rival that of HD networks, MAG networks might actually be preferred over HD networks in some cases. Because of their smaller number of generators and the relative ease with which they can be obtained, MAG can be a middle ground compromise between the basic Weyl generators and the difficult to find, but more efficient, HDs.

\section{Discussion} \label{sec:discussion}

In this work, we presented an analysis of the application of invariant algebras in machine learning, specifically in neural networks. We presented and extended some approximation theorems which utilise invariant algebras of polynomials to approximate invariant functions via neural networks. We then implemented and tested the performance of the network architectures of these theorems on polynomial regression under a variety of hyperparameters. We found an overall reduction in training loss and a more significant reduction in validation loss. These improvements were dependant on the maximal polynomial degree and the activation functions used. We also performed a Bayesian analysis of these networks and found that, beyond a minimum size, Hironaka decomposition networks always outperform minimal algebra generator networks and Weyl networks.

There are some drawbacks to this approach. The first is that computing generators of invariant algebras is hard in general. The silver-lining to this is that this costly procedure has to be done only once (and then one can just look up these generators from a database). Another drawback is that in some cases, there are just too many generators and one must build networks which are large enough to make use of their benefits and for the advantages to become apparent. 

Finally, there are a few avenues for future work on both the theoretical and computational sides. First, an interesting open problem to attack is the restriction of the Stone-Weierstrauss theorem to the case of group invariant functions. Although proved for complex-valued functions invariant under linearly reductive groups over $\C$ and real-valued functions invariant under compact groups over $\R$, the case of a general group is yet to be solved. This has important consequences even for systems which are only invariant under the Lorentz group, which is linearly reductive over $\R$ but not compact. Weyl's theorem for Lorentz-invariant polynomial generators is taken for granted when considering non-polynomial functions because of the assumption that we can approximate any continuous Lorentz-invariant function by the generators of the invariant polynomials. But as we have seen, this is not guaranteed.\footnote{For a counterexample where the generators of invariant polynomials cannot approximate the continuous invariant functions, consider the algebra in two variables $\R[x,y]^{\R^{\times}}$ invariant under the multiplicative group $\R^{\times}$ (which is not linearly reductive). It is easy to see that the only invariant polynomials are the constants $\R[x,y]^{\R^{\times}} \cong \R$. Now consider the function $f(x,y) = \frac{x}{y}$ defined on a compact subspace of the positive quadrant $\R_{+}^2$. This is continuous and invariant but cannot be approximated by the constant generators of $\R[x,y]^{\R^{\times}}$.} On the computational side, it would be interesting to see how these networks perform in different machine learning settings such as classification problems (in neural networks or otherwise) and unsupervised learning problems. Furthermore, it would be interesting to see whether the results found using the Bayesian approach still hold if likelihood-free inference methods are used to circumvent the need to choose a likelihood function. 

\section*{Acknowledgements}

WH would like to thank Gerald B. Folland, Ben Gripaios, Will Handley, Christopher G. Lester, and Joseph Tooby-Smith for enlightening discussions and helpful comments. Additionally, thanks to Will Handley for help with the \textsc{PolyChord} package and to Steve Wotton for help with the Cavendish HEP group HTC. WH is supported by the Cambridge Trust.

\appendix
\renewcommand*{\thesection}{\Alph{section}}
\section{Generalisations of the Stone-Weierstrauss theorem} \label{app:SW}

In this Appendix, we discuss a potential route to restricting the Stone-Weierstrauss theorem to the case of continuous functions invariant under some group $G$. That is, we would like to obtain a criterion on the group $G$ such that one could approximate any continuous $G$-invariant function by a $G$-invariant polynomial. As we have seen in Section \ref{sec:UAT}, in the case of linearly reductive groups over $\C$ and compact groups over $\R$, this is certainly possible. The problem with extending the approach used there to linearly reductive groups over $\R$ is that the maximal compact subgroups of such groups are not guaranteed to be Zariski-dense. 

For a different approach, we can reformulate the problem so that we may apply the non-invariant case of the Stone-Weierstrauss theorem to achieve our goal. We start by stating some definitions and theorems. Throughout this discussion, $X$ denotes a compact Hausdorff space (taken to be a compact subspace of $\C^n$) and $C(X)$ the $C^\star$ Banach algebra of continuous complex-valued functions, $f : X \rightarrow \C$, where the norm is defined to be the sup norm, $||f|| = \text{sup}_{x\in X} |f(x)|$.

\begin{definition}[Separating algebra]
We say that a subalgebra $A \subseteq C(X)$ separates $X$ if for any distinct $x,y \in X$, there exists an $f \in A$ such that $f(x) \neq f(y)$.
\end{definition}

Trivially, we have that $C(X)$ separates $X$.

\begin{theorem}[Stone-Weierstrauss theorem]
Let $A \subset C(X)$ be a subalgebra (closed under the unit and algebra multiplication operations on $C(X)$). A subalgebra inclusion $A \subset C(X)$ is dense if and only if it separates $X$. 
\end{theorem}

Now take $X$ to be a compact subspace of $\C^n$ and $A = \C[X]$ to be the polynomial algebra over $\C$ in the coordinates of $X$ in some basis. Consider a group $G$ acting on $X$, $G \times X \rightarrow X$, and denote by $C^G(X) \subset C(X)$ the $C^\star$ Banach subalgebra of continuous invariant complex-valued functions, $\{ f \in C(X) \ | \ f(g \circ x) = f(x) \ \forall g \in G \}$.

We would like to show that the subalgebra of invariant polynomials $\C[X]^G$ is dense in $C^G(X)$. To do this, we start by considering the orbit space $X/G$. Let $P : X \rightarrow X/G$ be the map that takes each $x \in X$ to its orbit. The topology on $X/G$ is the quotient topology. $X/G$ is furthermore compact (since it is the image of a compact space under a continuous map). We denote by $C(X/G)$ the $C^\star$ Banach algebra of continuous complex-valued functions, $f_G : X/G \rightarrow \C$. 

One can show that $C^G(X)$ is isomorphic to $C(X/G)$ (in the category $\mathbf{C^\star Alg}$). This allows us to reach our required result by applying the Stone-Weierstrauss theorem on $C(X/G)$. But care needs to be taken as, crucially, $X/G$ might not be Hausdorff in general (and even $C(X/G)$ could fail to separate the orbits). If it is, then the remainder of the problem is to show that the subalgebra of invariant polynomials separates the orbit space $X/G$. If not, then we can pass to a coarser quotient space as follows. Define two orbits $x,y \in X/G$ to be equivalent, $x \sim y$, if for all $f_G \in C(X/G)$, $f_G(x) = f_G(y)$. Denote the space of equivalence classes by $Y = X/G/ \sim$ giving it the quotient topology. One can then show that $C(Y)$ is isomorphic to $C^G(X)$ as well and furthermore, since it separates $Y$ (by definition), one can also show that $Y$ is Hausdorff. All that remains is to characterise the space of equivalence classes $Y$ and show that the subalgebra of invariant polynomials separates $Y$. 

There are two main bottlenecks to seeing this proof through to completion. First, it is difficult to check whether $Y$ is isomorphic to $X/G$. In other words, does $C(X/G)$ separate $X/G$? But say that one can establish which space is the correct one to consider. The second bottleneck is then showing that the subalgebra of invariant polynomials $\C[X]^G$ separates $X/G$ or $Y$. This is in general hard. Furthermore there are known examples where the subalgebra of invariant polynomials fails to separate the orbits (although in those cases, it is unknown whether $C^G(X)$ separates those orbits or not).

We hope that this is a motivating example, where a generalisation of the Stone-Weierstrauss theorem is necessary, that may encourage the expert (and interested) reader to investigate further.

\section{Nested sampling} \label{app:NS}

In this Appendix, we give a brief summary of nested sampling. Consider a model $M$ with parameters $\alpha$. Denote the likelihood of the data given the parameters by $L(\alpha) = P(D|\alpha, M)$ and the prior on these parameters by $\pi(\alpha) = P(\alpha|M)$. The posterior is then $P(\alpha | D, M) \propto L(\alpha)  \pi(\alpha)$.

The constant of proportionality of this relation is called the evidence, $Z(M)$. Computing $Z(M)$ amounts to evaluating the (usually) very high dimensional integral 
\begin{align*}
Z(M) = \int L(\alpha)  \pi(\alpha) d\alpha.
\end{align*}

The idea of nested sampling is to re-express this multi-dimensional integral as a one-dimensional one using a change of variables. Define $X(\lambda) = \int_{L(\alpha) > \lambda} \pi(\alpha) d\alpha$ to be the fraction of the prior (or prior volume) contained within an iso-likelihood contour $L(\alpha) = \lambda$. This is a decreasing positive semi-definite function. One may now write $Z(M)$, using a change of variable $L(\alpha)$ to $L(X)$, as 
\begin{align*}
Z(M) = \int L(X)  dX.
\end{align*}

Nested sampling evaluates this integral by sampling a set of $n_{\text{live}}$ {\em live points} from the prior, $\{ \alpha_i^{\text{live}} \}$, each with a likelihood $L_i = L(\alpha_i^{\text{live}})$ and $X_i = X(L_i)$. These points are then iteratively modified to converge exponentially onto the peaks of $L(X)$. The evidence can then be computed as a simple Riemann sum 
\begin{align*}
Z(M) = \sum_{i = 1}^{n_{\text{live}} - 1} (X_{i+1} - X_{i})L_i.
\end{align*}

For excellent overviews which deal with the many details that we have skimmed over, refer to the original nested sampling paper \cite{Skilling} and to the \textsc{PolyChord} package papers \cite{Polychord1, Polychord2}.

\section{Explicit generators} \label{app:gens}

In this Appendix, we provide an explicit list of generators for the example of $\C[V^4]^{SO(d) \times S_2 \times S_2}$ in Tables \ref{table:MAGgens} and \ref{table:HDgens}.

\def\arraystretch{1.5}%
\begin{table}[!htb]
\begin{adjustbox}{max width= \columnwidth}
\centering
\begin{tabular}{ |c|c| } 
\hline
\hline
& Minimal algebra generators \\
\hline
\hline
$g_1$  & $(p_1 \cdot p_1)+(p_2 \cdot p_2)$\\
\hline
$g_2$  & $(p_1 \cdot p_1) (p_2 \cdot p_2)$ \\
\hline
$g_3$  & $(p_3 \cdot p_3)+(p_4 \cdot p_4)$ \\
\hline
$g_4$  & $(p_3 \cdot p_3) (p_4 \cdot p_4)$ \\
\hline
$g_5$  & $(p_1 \cdot p_2)$  \\
\hline
$g_6$  & $(p_3 \cdot p_4)$  \\
\hline
$g_7$  & $(p_1 \cdot p_3)+(p_1 \cdot p_4) +(p_2 \cdot p_3)+(p_2 \cdot p_4)$ \\ 
\hline
$g_8$  & $(p_1 \cdot p_3) (p_1 \cdot p_4)+ (p_2 \cdot p_3) (p_2 \cdot p_4)$\\ 
\hline
$g_9$  & $(p_1 \cdot p_3) (p_2 \cdot p_3)+ (p_1 \cdot p_4) (p_2 \cdot p_4)$ \\ 
\hline
$g_{10}$  & $(p_1 \cdot p_4) (p_2 \cdot p_3)+ (p_1 \cdot p_3) (p_2 \cdot p_4)$  \\ 
\hline
$g_{11}$  & \makecell{$(p_1 \cdot p_1) (p_1 \cdot p_3)+(p_1 \cdot p_1) (p_1 \cdot p_4) +(p_2 \cdot p_2) (p_2 \cdot p_3)+(p_2 \cdot p_2) (p_2 \cdot p_4)$} \\ 
\hline
$g_{12}$  & \makecell{$(p_1 \cdot p_3) (p_3 \cdot p_3)+(p_2 \cdot p_3) (p_3 \cdot p_3)+(p_1 \cdot p_4) (p_4 \cdot p_4)+(p_2 \cdot p_4) (p_4 \cdot p_4)$}\\ 
\hline
$g_{13}$  & $(p_1 \cdot p_3)^3+(p_1 \cdot p_4)^3+(p_2 \cdot p_3)^3+(p_2 \cdot p_4)^3$ \\ 
\hline
$g_{14}$  & \makecell{$(p_1 \cdot p_1) (p_1 \cdot p_3)^2+(p_1 \cdot p_1) (p_1 \cdot p_4)^2+(p_2 \cdot p_2) (p_2 \cdot p_3)^2+(p_2 \cdot p_2) (p_2 \cdot p_4)^2$}\\ 
\hline
$g_{15}$  & \makecell{$(p_1 \cdot p_3)^2 (p_3 \cdot p_3)+(p_2 \cdot p_3)^2 (p_3 \cdot p_3)+(p_1 \cdot p_4)^2 (p_4 \cdot p_4)+(p_2 \cdot p_4)^2 (p_4 \cdot p_4)$} \\ 
\hline
$g_{16}$  & \makecell{$(p_1 \cdot p_1) (p_1 \cdot p_3) (p_3 \cdot p_3)+(p_2 \cdot p_2) (p_2 \cdot p_3) (p_3 \cdot p_3)+(p_1 \cdot p_1) (p_1 \cdot p_4) (p_4 \cdot p_4)+(p_2 \cdot p_2) (p_2 \cdot p_4) (p_4 \cdot p_4)$}  \\ 
\hline
$g_{17}$ & $\epsilon(p_1, p_2, p_3, p_4) ((p_1 \cdot p_3)-(p_1 \cdot p_4)-(p_2 \cdot p_3)+(p_2 \cdot p_4))$  \\
\hline
$g_{18}$ & $\epsilon(p_1, p_2, p_3, p_4) ((p_1 \cdot p_3)^2-(p_1 \cdot p_4)^2-(p_2 \cdot p_3)^2+(p_2 \cdot p_4)^2)$ \\
\hline
$g_{19}$ &  \makecell{$\epsilon(p_1, p_2, p_3, p_4) \big( (p_1 \cdot p_1) (p_1 \cdot p_3)- (p_1 \cdot p_1)(p_1 \cdot p_4) - (p_2 \cdot p_2)(p_2 \cdot p_3)+(p_2 \cdot p_2)(p_2 \cdot p_4) \big)$} \\
\hline
$g_{20}$ & \makecell{$\epsilon(p_1, p_2, p_3, p_4) \big( (p_1 \cdot p_3) (p_3 \cdot p_3)-(p_2 \cdot p_3) (p_3 \cdot p_3)- (p_4 \cdot p_4)(p_1 \cdot p_4)+ (p_4 \cdot p_4)(p_2 \cdot p_4) \big)$} \\
\hline
$g_{21}$ & $\epsilon(p_1, p_2, p_3, p_4) ((p_1 \cdot p_1)-(p_2 \cdot p_2)) ((p_3 \cdot p_3)-(p_4 \cdot p_4)) $ \\
\hline
\end{tabular}
\end{adjustbox}
\caption{\label{table:MAGgens} Table of a set of minimal algebra generators of $\C[V^4]^{SO(4) \times S_2 \times S_2}$}
\end{table}

\def\arraystretch{1.5}%
\begin{table}[!htb]
\centering
\begin{adjustbox}{max width= \columnwidth}
\begin{tabular}{ |c|c| } 
\hline
\hline
& Primaries \\ 
\hline
\hline
$\theta_1$  & $(p_1 \cdot p_1)+(p_2 \cdot p_2)$ \\
\hline
$\theta_2$  & $(p_1 \cdot p_1) (p_2 \cdot p_2)$ \\
\hline
$\theta_3$  & $(p_3 \cdot p_3)+(p_4 \cdot p_4)$  \\
\hline
$\theta_4$  & $(p_3 \cdot p_3) (p_4 \cdot p_4)$ \\
\hline
$\theta_5$  & $(p_1 \cdot p_2)$  \\
\hline
$\theta_6$  & $(p_3 \cdot p_4)$  \\ 
\hline
$\theta_7$  & $(p_1 \cdot p_3)+(p_1 \cdot p_4)+(p_2 \cdot p_3)+(p_2 \cdot p_4)$ \\
\hline
$\theta_8$  & $(p_1 \cdot p_3) (p_1 \cdot p_4)+(p_2 \cdot p_3) (p_2 \cdot p_4)$ \\
\hline
$\theta_9$  & $(p_1 \cdot p_3) (p_2 \cdot p_3)+(p_1 \cdot p_4) (p_2 \cdot p_4)$ \\
\hline
$\theta_{10}$  & $(p_1 \cdot p_4) (p_2 \cdot p_3)+(p_1 \cdot p_3) (p_2 \cdot p_4)$  \\
\hline
\hline
& Secondaries \\
\hline
\hline
$\eta_1$ & $1$ \\
\hline
$\eta_2$ & $ (p_1 \cdot p_1)(p_1 \cdot p_3)+(p_1 \cdot p_1)(p_1 \cdot p_4)+(p_2 \cdot p_2) (p_2 \cdot p_3)+ (p_2 \cdot p_2)(p_2 \cdot p_4)$ \\
\hline
 $\eta_3$ & $(p_1 \cdot p_3) (p_3 \cdot p_3)+(p_2 \cdot p_3) (p_3 \cdot p_3)+(p_1 \cdot p_4)(p_4 \cdot p_4) +(p_2 \cdot p_4) (p_4 \cdot p_4)$ \\
\hline
$\eta_4$ & $(p_1 \cdot p_3)^3+(p_1 \cdot p_4)^3+(p_2 \cdot p_3)^3+(p_2 \cdot p_4)^3$ \\
\hline
$\eta_5$&  $ (p_1 \cdot p_1)(p_1 \cdot p_3)^2+ (p_1 \cdot p_1)(p_1 \cdot p_4)^2+ (p_2 \cdot p_2)(p_2 \cdot p_3)^2+ (p_2 \cdot p_2)(p_2 \cdot p_4)^2$ \\
\hline
$\eta_6$& $(p_1 \cdot p_3)^2 (p_3 \cdot p_3)+(p_2 \cdot p_3)^2 (p_3 \cdot p_3)+(p_1 \cdot p_4)^2 (p_4 \cdot p_4)+(p_2 \cdot p_4)^2 (p_4 \cdot p_4)$ \\
\hline
$\eta_7$ & $(p_1 \cdot p_1) (p_1 \cdot p_3) (p_3 \cdot p_3)+(p_2 \cdot p_2) (p_2 \cdot p_3) (p_3 \cdot p_3)+(p_1 \cdot p_1) (p_1 \cdot p_4) (p_4 \cdot p_4)+(p_2 \cdot p_2) (p_2 \cdot p_4) (p_4 \cdot p_4)$\\ 
\hline
$\eta_8$ & $\epsilon(p_1,p_2,p_3,p_4) ((p_1 \cdot p_3)-(p_1 \cdot p_4)-(p_2 \cdot p_3)+(p_2 \cdot p_4))$\\ 
\hline
$\eta_9$ & $(p_1 \cdot p_1) (p_1 \cdot p_3)^2 (p_3 \cdot p_3)+(p_2 \cdot p_2) (p_2 \cdot p_3)^2 (p_3 \cdot p_3)+(p_1 \cdot p_1) (p_1 \cdot p_4)^2 (p_4 \cdot p_4)+(p_2 \cdot p_2) (p_2 \cdot p_4)^2 (p_4 \cdot p_4)$\\ 
\hline
$\eta_{10}$ & $\epsilon(p_1,p_2,p_3,p_4) ((p_1 \cdot p_3)^2-(p_1 \cdot p_4)^2-(p_2 \cdot p_3)^2+(p_2 \cdot p_4)^2)$\\ 
\hline
$\eta_{11}$  &$\epsilon(p_1,p_2,p_3,p_4) ((p_1 \cdot p_1) (p_1 \cdot p_3)- (p_1 \cdot p_1)(p_1 \cdot p_4) - (p_2 \cdot p_2)(p_2 \cdot p_3)+ (p_2 \cdot p_2)(p_2 \cdot p_4))$\\ 
\hline
$\eta_{12}$ & $\epsilon(p_1,p_2,p_3,p_4) ((p_1 \cdot p_3) (p_3 \cdot p_3)-(p_2 \cdot p_3) (p_3 \cdot p_3)-(p_1 \cdot p_4) (p_4 \cdot p_4) +(p_2 \cdot p_4) (p_4 \cdot p_4))$\\ 
\hline
$\eta_{13}$  & $\epsilon(p_1,p_2,p_3,p_4) ((p_1 \cdot p_1)-(p_2 \cdot p_2)) ((p_3 \cdot p_3)-(p_4 \cdot p_4))$\\ 
\hline
$\eta_{14}$  & $\epsilon(p_1,p_2,p_3,p_4) ((p_1 \cdot p_1) (p_1 \cdot p_3)^2- (p_1 \cdot p_1)(p_1 \cdot p_4)^2 -(p_2 \cdot p_2)(p_2 \cdot p_3)^2+ (p_2 \cdot p_2)(p_2 \cdot p_4)^2)$\\ 
\hline
$\eta_{15}$ & $\epsilon(p_1,p_2,p_3,p_4) ( (p_3 \cdot p_3)(p_1 \cdot p_3)^2- (p_3 \cdot p_3)(p_2 \cdot p_3)^2 - (p_4 \cdot p_4)(p_1 \cdot p_4)^2+ (p_4 \cdot p_4)(p_2 \cdot p_4)^2)$\\ 
\hline
$\eta_{16}$  & $\epsilon(p_1,p_2,p_3,p_4) \big( (p_1 \cdot p_1) (p_3 \cdot p_3) (p_1 \cdot p_3)^3 -(p_2 \cdot p_2) (p_3 \cdot p_3)(p_2 \cdot p_3)^3$  \\ 
& $-(p_1 \cdot p_1) (p_4 \cdot p_4) (p_1 \cdot p_4)^3+(p_2 \cdot p_2) (p_4 \cdot p_4) (p_2 \cdot p_4)^3 \big)$ \\
\hline
\end{tabular}
\end{adjustbox}
\caption{\label{table:HDgens} Table of the primary and secondary generators of a Hironaka decomposition of $\C[V^4]^{SO(4) \times S_2 \times S_2}$}
\end{table}

\section*{References}

\bibliographystyle{elsarticle-num}
\bibliography{IPML}

\begin{thebibliography}{10}
\expandafter\ifx\csname url\endcsname\relax
  \def\url#1{\texttt{#1}}\fi
\expandafter\ifx\csname urlprefix\endcsname\relax\def\urlprefix{URL }\fi
\expandafter\ifx\csname href\endcsname\relax
  \def\href#1#2{#2} \def\path#1{#1}\fi

\bibitem{LPIP}
B.~Gripaios, W.~Haddadin, C.~G. Lester, Lorentz- and permutation-invariants of
  particles, Journal of Physics A: Mathematical and Theoretical 54~(15) (2021)
  155201.
\newblock \href {http://dx.doi.org/10.1088/1751-8121/abe58c}
  {\path{doi:10.1088/1751-8121/abe58c}}.

\bibitem{Lorentz-Top}
A.~Bogatskiy, B.~Anderson, J.~T. Offermann, M.~Roussi, D.~W. Miller, R.~Kondor,
  {Lorentz Group Equivariant Neural Network for Particle Physics}\href
  {http://arxiv.org/abs/2006.04780} {\path{arXiv:2006.04780}}.

\bibitem{dk}
H.~Derksen, G.~Kemper, {Computational invariant theory}, Springer,
  Berlin/Heidelberg, DE, 2002.

\bibitem{Weyl}
H.~Weyl, {The classical groups: their invariants and representations},
  Princeton University Press, Princeton, NJ, US, 1966.

\bibitem{SW1}
M.~H. Stone, The generalized weierstrass approximation theorem, Mathematics
  Magazine 21~(4) (1948) 167--184.

\bibitem{SW2}
M.~H. Stone, Applications of the theory of boolean rings to general topology,
  Transactions of the American Mathematical Society 41~(3) (1937) 375--481.

\bibitem{Yarotsky}
D.~Yarotsky, Universal approximations of invariant maps by neural networks
  (2018).
\newblock \href {http://arxiv.org/abs/1804.10306} {\path{arXiv:1804.10306}}.

\bibitem{Kraft}
H.~Kraft, C.~Procesi, Classical invariant theory: a primer, 1996.

\bibitem{Pinkus}
A.~Pinkus, Approximation theory of the mlp model in neural networks, Acta
  Numerica 8 (1999) 143–195.
\newblock \href {http://dx.doi.org/10.1017/S0962492900002919}
  {\path{doi:10.1017/S0962492900002919}}.

\bibitem{Voigtlaender}
F.~{Voigtlaender}, {The universal approximation theorem for complex-valued
  neural networks}, arXiv e-prints.

\bibitem{Poly-nets1}
K.~F.~E. Chong, A closer look at the approximation capabilities of neural
  networks (2020).
\newblock \href {http://arxiv.org/abs/2002.06505} {\path{arXiv:2002.06505}}.

\bibitem{Poly-nets2}
A.~Andoni, R.~Panigrahy, G.~Valiant, L.~Zhang, Learning polynomials with neural
  networks 32~(2) (2014) 1908--1916.

\bibitem{NNmetric}
A.~Butter, G.~Kasieczka, T.~Plehn, M.~Russell, Deep-learned top tagging with a
  lorentz layer, {SciPost} Physics 5~(3) (2018) 28.
\newblock \href {http://dx.doi.org/10.21468/scipostphys.5.3.028}
  {\path{doi:10.21468/scipostphys.5.3.028}}.

\bibitem{HandleyBNN}
K.~Javid, W.~Handley, M.~Hobson, A.~Lasenby, Compromise-free bayesian neural
  networks (2020).
\newblock \href {http://arxiv.org/abs/2004.12211} {\path{arXiv:2004.12211}}.

\bibitem{Like-free}
X.~Didelot, R.~G. Everitt, A.~M. Johansen, D.~J. Lawson, Likelihood-free
  estimation of model evidence, Bayesian Analysis 6~(1).
\newblock \href {http://dx.doi.org/10.1214/11-ba602}
  {\path{doi:10.1214/11-ba602}}.

\bibitem{Polychord1}
W.~J. Handley, M.~P. Hobson, A.~N. Lasenby, polychord: nested sampling for
  cosmology, Monthly Notices of the Royal Astronomical Society: Letters 450~(1)
  (2015) L61–L65.
\newblock \href {http://dx.doi.org/10.1093/mnrasl/slv047}
  {\path{doi:10.1093/mnrasl/slv047}}.

\bibitem{Polychord2}
W.~J. Handley, M.~P. Hobson, A.~N. Lasenby, polychord: next-generation nested
  sampling, Monthly Notices of the Royal Astronomical Society 453~(4) (2015)
  4385–4399.
\newblock \href {http://dx.doi.org/10.1093/mnras/stv1911}
  {\path{doi:10.1093/mnras/stv1911}}.

\bibitem{Mackay}
D.~J.~C. MacKay, A practical bayesian framework for backpropagation networks,
  Neural Comput. 4~(3) (1992) 448–472.
\newblock \href {http://dx.doi.org/10.1162/neco.1992.4.3.448}
  {\path{doi:10.1162/neco.1992.4.3.448}}.

\bibitem{Skilling}
J.~Skilling, Nested sampling for general bayesian computation, Bayesian
  Analysis 1~(4).
\newblock \href {http://dx.doi.org/10.1214/06-ba127}
  {\path{doi:10.1214/06-ba127}}.

\end{thebibliography}

\end{document}